\documentclass[12pt]{article}
\usepackage{graphicx}

\title{\bf The impact of the Sakata model}
\author{L.B. Okun \\
{\it Institute of Theoretical and Experimental Physics} \\ {\it
Moscow, Russia}}
\date{}

\begin{document}

\maketitle

\begin{abstract}

The evolution of the Sakata model is described on the basis of
personal recollections, proceedings of international conferences
on high energy physics and some journal articles.
\end{abstract}

\section{1956. Sakata at ITEP}

Shoichi Sakata was the first foreigner who visited the ITEP theory
division. He came in the spring of 1956 and compiled a list of the
ITEP theorists -- I.Ya. Pomeranchuk, V.B. Berestetsky, A.D.
Galanin, A.P. Rudik, B.L. Ioffe, V.V. Sudakov,  I.Yu. Kobzarev and
myself. Sakata also took a photo of those who were present. (It
would be interesting to find this picture in his archives.) I
still have the three pages of thin rice paper with the Sakata
model which he left with us. They correspond to his paper \cite{1}
. These three pages were crucial for all my life in physics.

Sakata \cite{1} considered 7 mesons (3 $\pi$, 4 $K$) and 8 baryons
(2 $N$, $\Lambda$, 3 $\Sigma$, 2~$\Xi$) known at that time. He
postulated that 3 baryons -- $p, n, \Lambda$ -- are more
fundamental than the other 5 baryons and 7 mesons and demonstrated
that these 12 particles could be composed from $p, n, \Lambda$ and
$\bar p, \bar n, \bar\Lambda$. The paper had a philosophical
flavor and contained no experimental predictions. In 1956 particle
physicists were discussing the $\tau\theta$-puzzle and parity
violation (see reference \cite{1d} for further details). Therefore
the paper \cite{1} as well as three accompanying papers of
Sakata's students \cite{1a,1b,1c} had no immediate response.
(S.~Tanaka \cite{1b} discussed $\tau\theta$-parity degeneracy in
the Sakata model, Z.~Maki \cite{1c} attempted to calculate bound
states of baryons and antibaryons, while K.~Matumoto \cite{1a}
suggested a semi-empirical formula for masses of composite
particles.)

\section{1957. Padua -- Venice}

In the summer of 1957 I suddenly ``reinvented'' the Sakata model
and realized its beauty and its potential. Then I recalled the
three rice pages and reread them.

My first paper \cite{2} on the Sakata model was presented by
I.I.~Gurevich at the conference in Padua -- Venice, September
1957. A slightly different text \cite{3} was published in a
Russian journal. In these publications the three ``sakatons'' were
not physical $p,n,\Lambda$, but some primary particles denoted by
the same letters, so ``we can assume that for the primary
particles $m_\Lambda = m_N$'' \cite{3}. Strong and weak
interactions of sakatons were considered and for the latter a
number of selection rules were deduced, in particular, those which
are known as $|\Delta S|=1$, $\Delta T=1/2$ for nonleptonic decays
of strange particles via the $\bar n\Lambda$ transition, while for
the leptonic (or semi-leptonic) ones $|\Delta S|=1$, $\Delta
Q=\Delta S$ and $\Delta T=1/2$ via $\bar p\Lambda$ current.

As for the strong interactions, the existence of $\eta$- and
$\eta^\prime$-mesons was predicted in \cite{2,3}; I denoted them
$\rho_1^0$ and $\rho_2^0$:

``In the framework of this scheme there is a possibility of two
additional neutral mesons which have not so far been observed: $$
\rho_1^0 = \Lambda \bar\Lambda \; , \;\; \rho_2^0 = (p\bar p -
n\bar n)/\sqrt 2 \;\; . $$

The isotopic spin of the $\rho$-mesons is zero.'' \cite{2}

(The unconventional minus sign in the definition of $\rho_2^0$ was
in accord with the not less unconventional definition $\pi^0 =
(p\bar p  +n\bar n)/\sqrt 2$.)

\section{1957. Stanford and Berkeley}

In December 1957 four Soviet particle physicists
(D.I.~Blokhintsev, V.P.~Dzhelepov, S.Ya.~Nikitin and myself)
visited Palo Alto, Berkeley, Boston, New York, Brookhaven. For me
it was my first trip abroad and the first flight in my life. (The
next time the Soviet authorities allowed me to visit the USA was
only in 1988 for the Neutrino'88 conference.)

During the 1957 trip I talked with M.~Baker, H.~Bethe, S.~Drell,
R.~Feynman, R.~Gatto, M.~Gell-Mann, S.~Goldhaber, C.~Sommerfield,
F.~Zachariasen, C.~Zemach and many others, gave a seminar at
Berkeley. As a result of this seminar E.~Segre invited me to write
an article for the Annual Review of Nuclear Science. It appeared
in 1959 (see below).

\section{1958. Geneva}

My second paper on the Sakata model ``Mass reversal and compound
model of elementary particles'' was published in June 1958 as a
Dubna preprint \cite{5} and I had it with me at the 1958 Rochester
Conference at CERN. On the initiative of J.R.~Oppenheimer and
R.E.~Marshak a special seminar was arranged at which I presented
my paper at the start of the conference and then was asked to
present it also at Session 7, ``Special theoretical topics'', see
\cite{6}. (Note that selection rules for weak interactions in
sections 14, 15 and references 24-28 of the Dubna preprint
\cite{5} were deleted by the editors of the Proceedings  \cite{6};
see the Appendix for the deleted pages.)

In \cite{5,6} $\rho_1^0$ and $\rho_2^0$ became mixtures of the
states discussed above. What is more important, all interactions
were assumed to be $\gamma_5$-invariant following  papers
\cite{9a,10a,11a} and especially \cite{10b}. The conservation of
the vector non-strange current, postulated in \cite{6a,9a}, was
shown in \cite{5,6} to be inevitable in the Sakata model.
Unfortunately the strong interaction was written  as an ugly
four-fermion interaction of sakatons.

The discussion of my talk involved R.~Gatto, G.~L\"{u}ders,
R.~Adair, G.~Wentzel, T.D.~Lee, Y.~Yamaguchi (see page 228 of the
Proceedings). The discussion with Yoshio Yamaguchi continued
during the lunch in the CERN canteen. In the afternoon of the same
day J.~Oppenheimer commented my argument that in the Sakata model
conservation of the weak non-strange vector current is inevitable
(see page 257).  He again at length commented the subject in his
``Concluding Remarks'' at the Conference (see page 293).
R.~Marshak stressed the novelty of chiral invariance for strong
interactions (see page 257). In his talk ``$K_{e3}$ and $K_{\mu
3}$ decays and related subjects'' Marshak repeatedly underlined
that for these decays ``$\Delta I=1/2$ in Okun's model'' (see
\cite{7}, pp. 284, 285).

In the discussion \cite{14a} I described an upper limit on $\Delta
S =2$ transitions which had been derived by B.~Pontecorvo and
myself \cite{9b}.

On the basis of the selection rules for weak interactions which
follow from the Sakata model the lifetime of $K_2^0$  and its
branching ratios were predicted \cite{8} by I.Yu.~Kobzarev and
myself. This prediction was cited by me in December 1957 at
Stanford and as reference [28] in the Dubna preprint \cite{5} and
was soon confirmed experimentally \cite{9}.

\section{1959. Kiev symmetry}

In 1959 my paper \cite{4} appeared as well as its Russian twin
\cite{4a}. I received a hundred requests for reprints, many of
them -- from Japan. Strangely enough, rereading now this paper, I
do not see in it the prediction of $\eta$ and $\eta^\prime$ and
any statement that $p, n, \Lambda$ are not physical baryons, but
some more fundamental particles. Both the prediction and the
statement were in \cite{2,3,5}. I cannot understand now their
irrational omission in \cite{4,4a}.

In 1959 other authors started to publish papers on the Sakata
model. A.~Gamba, R.~Marshak and S.~Okubo \cite{10} pointed out the
symmetry between the three leptons ($\mu,e,\nu$) and three baryons
($\Lambda, n,p$) ``in models of Sakata \cite{1} and Okun
\cite{3}''\footnote{Here and in other quotations the reference
numbers correspond to my list of references.}. This symmetry has
been emphasized by Marshak (in his rapporteur talk \cite{11} at
the 1959 Rochester conference in Kiev) and became known as the
Kiev symmetry. (I served as a scientific secretary of R.~Marshak
and participated in preparation of his report.)

At the Kiev conference M.~Gell-Mann told me: ``If I were you, I
would introduce in the $\Lambda np$ model the linear superposition
($n\cos\theta + \Lambda \sin\theta$)''. I do not understand why I
did not follow his advice. The angle $\theta$ is known now as the
Cabibbo angle. The weak current $\bar p (n + \varepsilon
\Lambda)/(1+ \varepsilon^2)^{1/2}$ first appeared  next year in
the paper by M.~Gell-Mann and M.~Levy \cite{13a}.

In 1959 the symmetry which is now called SU(3) was introduced into
Sakata model. Y.~Yamaguchi \cite{12} with reference to \cite{6}
stressed the existence of 9 pseudoscalar mesons ($9 = \bar 3
\times 3$). O.~Klein \cite{23a} and  S.~Ogawa \cite{13} discussed
generalizations of isotopic symmetry. In particular, S.~Ogawa with
reference to \cite{12} considered 3 doublets $(pn)$, $(n\Lambda)$,
$(\Lambda p)$ and 3 meson triplets. M.~Ikeda, S.~Ogawa, Y.~Ohnuki
\cite{14} with reference to \cite{13} developed some mathematical
constructs of the symmetry to which they referred as U(3). O.
Klein \cite{23a} discussed the interaction between the triplet of
sakatons and the octet of pseudoscalar mesons and stressed the
symmetry between $\Lambda n p$ and $\mu e \nu$.

\section{1960. Rochester}

In 1960 I was invited to give a rapporteur talk at the Rochester
Conference in Rochester. I prepared the draft of the talk, but was
not allowed by Soviet authorities to attend the conference. My
draft \cite{15} based on the Sakata model has been prepared for
the Proceedings by S.~Weinberg.  M.L.~Goldberger who ``was thrown
into a breach at a rather late date'' served as a rapporteur on
``Weak interactions (theoretical)''  \cite{16} referred to my
draft. R.~Feynman  \cite{17} spoke on the conserved vector
current. He said that in the model of Fermi and Yang ``as has been
pointed out in much more detail by Okun, in any complex structure,
the coupling of the beta decay is proportional to the total
isotopic spin''. M.~Gell-Mann \cite{18} spoke on the conserved and
partially conserved currents . He said: ``...there is the scheme
mentioned by Feynman and favored by Okun, Marshak, and others,
based on just $n,p$, and $\Lambda$. Of course, if that is right we
do not need the elaborate machinery I just described. We simply
draw an analogy''. But as it is clear from their talks both
Feynman and Gell-Mann at that time preferred to use the composite
model only as a tool to formulate more general phenomenological
approaches. Among the talks at Rochester 1960 was that by
Y.~Ohnuki \cite{19} who with a reference to \cite{3} assumed
$m_\Lambda =m_N$ and the three-dimensional unitary symmetry.

An important paper of 1960 was that by J.~Sakurai \cite{20}. With
a reference to \cite{6} he mentioned that instead of $N,\Lambda$
one can use as ``elementary'' $\Xi,\Lambda$. He considered the
absence of $\eta$-meson as a serious problem: ``...within the
framework  of Fermi--Yang--Sakata--Okun model it may be difficult
to explain why the $\eta$ does not exist'' (see pp. 32-36).

The $\eta$-meson was discovered within a year \cite{21} .

Further progress in SU(3) symmetric Sakata model was achieved by
M.~Ikeda, Y.~Miyachi, S.~Ogawa \cite{36b}, who applied this
symmetry to weak decays. Z.~Maki, M.~Nakagava, Y.~Ohnuki,
S.~Sakata published a paper on Sakata model \cite{24}. They wrote:
``... it has recently become clear that Feynman--Gell-Mann current
derived from the Sakata model is quite sufficient to account for
the experimental facts concerning the weak processes
\cite{25,3}''. They postulated the existence of a so-called $B^+$
matter. The bound state $e B^+$ had been identified with $n$,
bound state $\mu B^+$ -- with $\Lambda$, while $\nu B^+$ -- with
$p$.

In 1960--61 I was giving lectures \cite{23aa,23b} based on the
Sakata model. Subsequently they were recast into the book
\cite{23c}. My major mistake at that time was that I did not
consider seriously eight spin 1/2 baryons as an SU(3) octet in
spite of the ``eightfold way'' papers by M.~Gell-Mann \cite{38a,
39b} and Y.~ Ne'eman \cite{38b}. (The former referred to papers
\cite{14,12,4}.)

\section{1962. Geneva again}

In 1962 the Sakata model was ``falsified'' for a short time by
experiments \cite{29,30}, which discovered decays $\Sigma^+\to
n\mu^+\nu$ and $K^0\to e^+ \nu\pi^-$ forbidden by $\Delta S =
\Delta Q$ rule. At the 1962 Geneva conference I tried to find a
mistake in the results \cite{29,30} but failed. Pomeranchuk who
witnessed the argument commented later that my ``feathers were
flying''. I do not remember now how the mistake was found
subsequently by experimentalists. Maybe it was a statistical
fluctuation.

The authors of articles \cite{29,30} referred to the paper by
Feynman and Gell-Mann \cite{9a}. While in my papers \cite{2,3} the
forbidden decays were simply listed, in \cite{9a} the notations
$\Delta Q$ and $\Delta S$ were used and the currents with $\Delta
Q = \Delta S$ and $\Delta Q = -\Delta S$ ($\bar p \Lambda$ and
$\bar n \Sigma^+$-currents) were phenomenologically considered on
the same footing. The product of these currents gives transitions
with $\Delta S =2$. The limit on these transitions from the
absence of decays $\Xi^- \to n\pi^-$ was not reliable because ``so
few $\Xi$ particles have been seen that this is not really
conclusive''\cite{9a}. (The paper \cite{9b} (published in June
1957) had put a much better limit on $\Delta S =2$ processes from
$K^0\leftrightarrow \bar K^0$ transitions. But it was not known to
Feynman and Gell-Mann when they wrote \cite{9a}.)

In 1962 M.~Gell-Mann predicted the existence of $\Omega$-hyperon
\cite{39a}. I.~Kobzarev and myself \cite{39c} derived the SU(3)
relations between semileptonic decays of $\pi$ and $K$-mesons.
Together with relations for the decays of baryons they were later
derived by N.~Cabibbo \cite{39d}.

\section{1962. From 3 to 4 sakatons}

The discovery of $\nu_\mu$ prompted attempts  to reconcile the
existence of two neutrinos with the lepton-sakaton symmetry. In
order to preserve the Kiev symmetry Z.~Maki, M.~Makagawa,
S.~Sakata \cite{27} modified the $B^+$ matter model  \cite{24}.
They assumed that $p=\nu_1, B^+$, where $\nu_1$ is one of the two
orthogonal superpositions of $\nu_e$ and $\nu_\mu$. The other
superposition $\nu_2$ was assumed either not to form at all a
bound state with $B^+$ or to form a baryon with a very large mass.
On the basis of this model the paper introduced $\nu_e - \nu_\mu$
oscillations. Another way to lepton-sakaton symmetry was suggested
in the paper by Y.~Katayama, K.~Motumoto, S.~Tanaka, E.~Yamada
\cite{46a}, where the fourth sakaton was explicitly introduced.

\section{1964. Quarks}

In 1964 $\eta^\prime$-meson and $\Omega$-hyperon were discovered
\cite{22,43a}. Earlier this year G.~Zweig \cite{Y} and M.
Gell-Mann \cite{X} replaced the integer charged sakatons by
fractionally charged particles (aces -- Zweig; quarks --
Gell-Mann). This allowed them to construct not only the octet and
singlet of mesons, but also the octet and decuplet  of baryons.
When establishing  the electromagnetic and weak currents in the
quark model M.~Gell-Mann \cite{X} referred to similar expressions
in the Sakata model.

\section{November 2006 and afterwards}

On November 3 2006 I received the following email from a colleague
and a friend of mine -- Valentine I. Zakharov:

``Dear Lev Borisovich,

I am now visiting Kanazawa, Japan. This month, there will be a
one-day conference in Nagoya, to celebrate 50 years of the Sakata
model. They invited me to come and I eagerly agreed.

One of the reasons -- which you can readily guess-- was that the
words `Sakata model' were among the first ones I heard about our
field. (You were giving lectures to `experimentalists', with
Alikhanov in the first row; (M.I. Ryazanov from MEPHI encouraged
me to attend; it was some time before I showed up later).

I will mention of course that you were developing the Sakata model
at ITEP. But, unfortunately, I realized that I do not know
anything else, to any extent personal about Sakata-sensei. I mean,
no other papers, or their echo in Russia/USSR, nothing ... May be
you can help in some way?

Excuse me, please, for bothering you and with best regards,
Valya''.

To answer Valentine's request I have written this brief review.
Thinking that it might be of more general interest, I published it
as version 1 of hep-ph/0611298.

On December 22 I received an email from K.~Yamawaki who kindly
invited me to publish this paper in the Proceedings of the Sakata
Model Symposium. In editing the paper I benefited from email
exchanges with S.~Pakvasa, H.~Lipkin and A.~Gal. Another
interpretation of the terms ``Sakata model'' and ``Sakata
symmetry'' one can find in preprints \cite{52,53}.

\newpage

\section*{Appendix}

Four pages from the Dubna preprint \cite{5}:

\begin{center}

\bigskip

\includegraphics[width=0.9\textwidth]{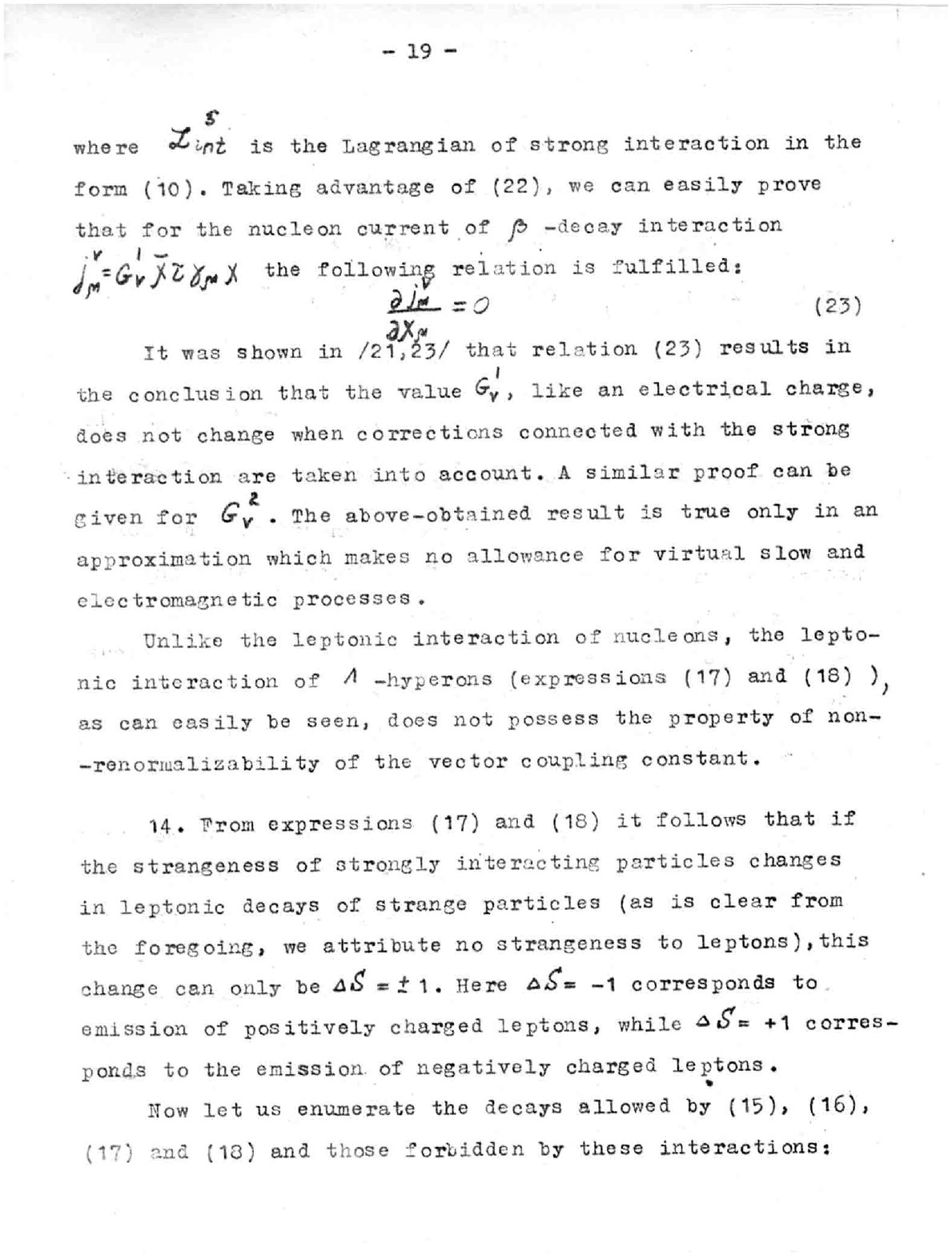}

\includegraphics[width=1.0\textwidth]{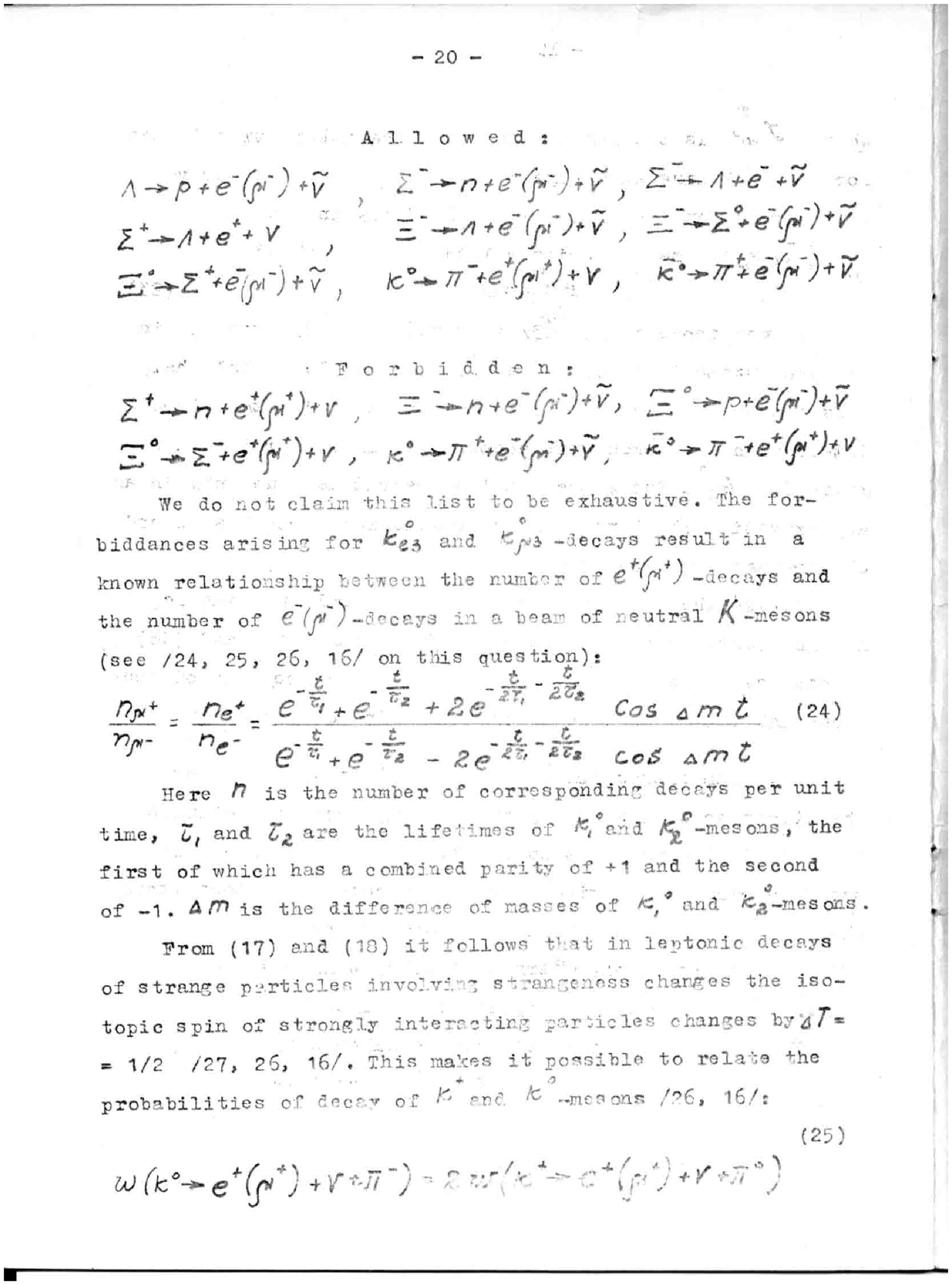}

\includegraphics[width=1.0\textwidth]{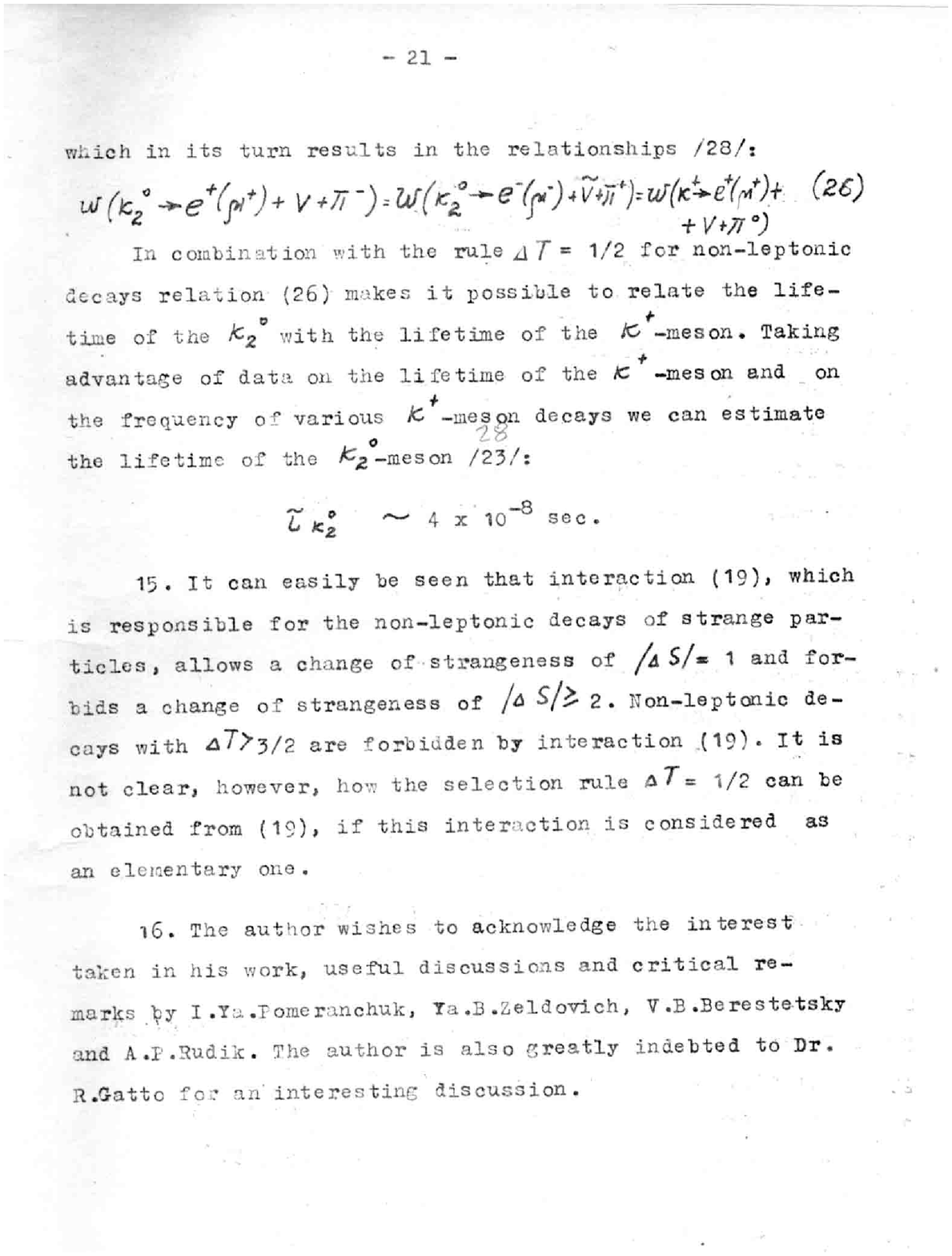}

\includegraphics[width=1.0\textwidth]{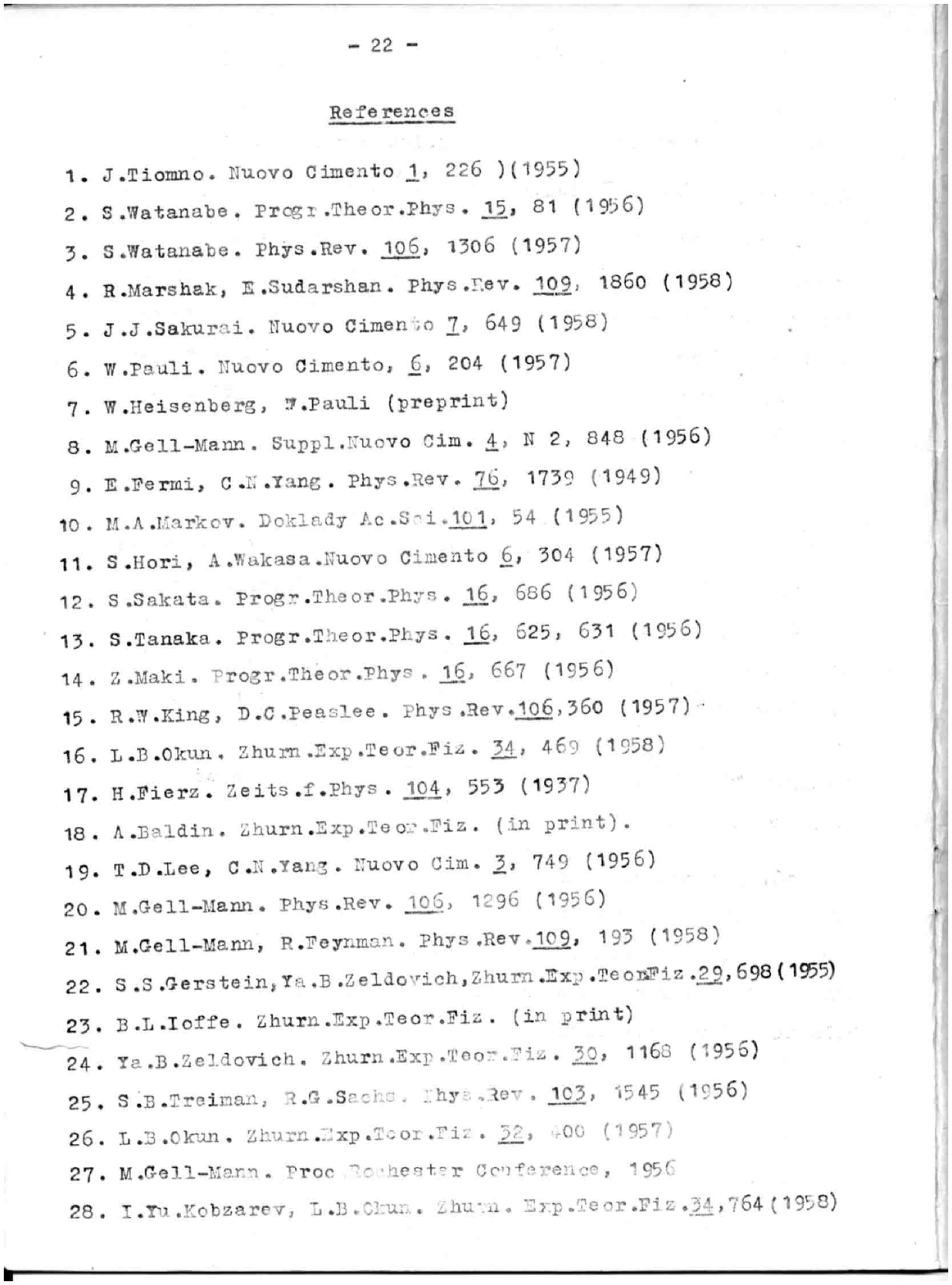}

\end{center}

\end{document}